\documentstyle[prl,aps,epsf,preprint]{revtex}
\begin{document}
\draft
\title {Atomistic Description of Shallow Levels in Semiconductors}
\author{A. S. Martins, J. G. Menchero, R. B. Capaz, and Belita Koiller}
\address{Instituto de F\'\i sica, Universidade Federal do Rio de Janeiro, 
Cx.P. 68528, Rio de Janeiro, RJ 21945-970, Brazil}
\date{\today}
\maketitle

\begin{abstract}
The wave function and binding energy for shallow donors in GaAs are calculated 
within the tight binding (TB) approach, for supercells containing up to two 
million atoms. The resulting solutions, coupled with a scaling law, allow 
extrapolation to the bulk limit. A sharp shallow-deep transition is obtained as 
the impurity perturbation increases. The model allows investigating the 
quantitative consistency between the effective mass theory and the TB 
formalism. Although the calculated binding energies are in excellent 
agreement,  anisotropies and the overall decay obtained in the TB envelope 
function can not be afforded by the hydrogenlike effective mass prediction.

\end{abstract}
\pacs{PACS numbers: 71.15.Ap, 71.55.-i, 71.55.Eq}

Electronic states in condensed matter systems may exist in either localized 
atomic-like states or in delocalized running-wave states \cite{anderson}. Such 
dualistic behavior has led naturally to two well-established frameworks in 
which to describe the electronic behavior of solids: 
Real space pictures privilege the atomic aspects of electronic behavior, 
whereas Bloch states provide a foundation to describe the wave-like properties. 
For an ideal periodic crystal, these two views were reconciled several decades 
ago, when Wannier \cite{wannier} showed that a set of orthogonal ``atomic'' 
functions localized at the crystal's atomic sites may be obtained from the 
Bloch eigenfunctions, and that either may be taken as the basis set for the 
general description of electrons in solids. Any electronic theory which adopts 
a basis set localized in real space is generally referred to as following the 
tight-binding (TB) approach. In some practical cases, it is also possible to 
utilize a basis set localized in reciprocal space about an appropriate point in 
the Brillouin zone. Such an approach then gives rise to the so-called Effective 
Mass Theory (EMT) \cite{slater}. EMT is the most successful comprehensive 
approach for the description of electronic excitations in insulators, of 
transport and optical properties in semiconductors, of dynamics of electrons in 
metals \cite{ziman}.

A crowning achievement, dating as far back as the 1950's, of the EMT is 
undoubtedly the quantitative understanding obtained of shallow impurity states 
in semiconductors \cite{emt1,emt2,pantelides}. A complementary TB theory for 
{\em deep} impurities was presented later by Hjalmarson {\it et al}, \cite
{HVWD}. The choice of approach in each case is defined naturally, given the 
extended nature of the shallow levels described within EMT versus the localized 
nature of the deep states considered in the TB scheme. 
A basic limitation of 
the EMT, however, is that the approximation (i.e., a single dominant k-point) 
breaks down as the impurity state becomes increasingly delocalized throughout 
the Brillouin zone, such as the case 
for deep levels. Similarly, a practical limitation of the TB approach is that 
until now it could
not be extended to the shallow impurity limit due to the slow decay of the 
defect wave function which requires extremely large supercells for meaningful 
calculations. As a result, for decades a gap has existed between two of the 
most widely used theories for impurities in semiconductors: The TB approach was 
confined to deep states, whereas the EMT was used for shallow states.  

In the present work we demonstrate that the gap between shallow and deep levels may 
be bridged via the TB formalism in conjunction with extrapolation schemes for the
binding energies \cite{form1} and for the envelope functions. Our proposed atomistic 
description provides a flexible way to explore and test key predictions of EMT for shallow levels. In particular, the following points are addressed and explained: (a) The values of ionization energies, (b) the independence of these ionization energies on the impurity 
species, and (c) the shape of the envelope function for the defect state. We 
show that wave functions may substantially deviate from the hydrogen-like 
predictions of EMT. Besides the obvious academic interest of unifying shallow 
and deep impurity level descriptions within a single theoretical approach, 
problems of current interest in nanostructured materials demand an atomistic 
description for which the TB approach coupled to the theoretical tools adopted 
herein
may provide a reliable alternative to EMT and to {\it ab-initio} treatments 
based in the local density approximation. 

In the EMT, it is assumed that the impurity wave function 
is highly delocalized in real space, which in turn implies a strong 
localization  in \textit{k}-space. 
The electronic properties of the host material are thus described by a few 
parameters related to the energy bands near the set of \textit{k}-points in 
consideration. In particular, effective masses are defined from the bands' 
curvatures at these points. The simplest case is that of a nondegenerate 
conduction band with an extremum at $\Gamma $ $({\bf k}=0)$, corresponding to 
the Bloch eigenstate $\psi_C^{\Gamma} ({\bf r})$ and eigenvalue $E_C (\Gamma)$. 
The cubic symmetry of the lattice implies that the nondegenerate band is 
isotropic around this extremum, leading to a single effective mass $m^*$: $E_C
({\bf k}\approx 0) = E_C(\Gamma) + \hbar^2 {\bf k}^2/2m^*$.  
Following the original EMT formulation \cite{emt2}, the perturbation potential 
for donor impurities is
\begin{equation}
 U(r)=-{e^{2}}/{\varepsilon r},
 \label{pot}
 \end{equation}
where $\varepsilon$ is the static dielectric constant of the host. 
Within a set of additional approximations \cite{pantelides}, 
the impurity wave function is written as $\Psi({\bf r}) = F({\bf
r})\psi_C^\Gamma ( {\bf r})$, 
and the eigenvalue problem for the one-electron Hamiltonian  $H=H_0+U(r)$,
where
$H_0$ describes the perfect host material, maps into an 
hydrogenic equation 
\begin{equation}
[-\hbar^2\bigtriangledown^2/2m^*  + U(r) ]F(\textbf{r}) = [E-E_C(\Gamma)]
F(\textbf{r})
\label{diferential}
\end{equation}
leading to the ground state envelope function 
$ F(\textbf{r})= (1/\sqrt{\pi {a^*}^3})\exp(-r/a^*)$, 
 where $a^*=a_0\varepsilon({m_0/m^*})$ is the effective Bohr radius, $a_0$=0.53 
\AA~ is the Bohr radius of the hydrogen atom and $m_0$ is the
free electron mass. The EMT basic assumptions are fulfilled if $a^*$ is much 
larger than the
host lattice parameter. In this case, the donor binding energy for a single 
impurity is
$ E_{d}^{*}=E_H\,[m^*/(\varepsilon^{2}{m_0})]$, where $E_H$=13.6 eV is the 
hydrogen ionization energy.

For donor levels in semiconductors with a conduction-band minimum at $\Gamma$, 
the above model combines simplicity and physical intuition and moreover leads 
to a good agreement with experimental results. In the case of GaAs, 
$\varepsilon=12.56$ and $m^*/m_0=0.068$, leading to
$a^*=97.7$ \AA $\;$ (which is indeed much larger than the host lattice 
parameter ${\rm
a_{GaAs}}$ = 5.65\AA), and to $E_{d}^{*}=5.86$ meV, which is in remarkable 
agreement with the
experimental data for donor binding energies in this material \cite{exper}. It 
is intriguing that most donor levels in GaAs are essentially independent of the 
substitutional species, e.g. 5.91, 5.84 and 5.88 meV respectively for C, Si and 
Ge replacing Ga, while acceptor levels vary 
by almost a factor of two for these same species substituting As \cite{form1}. 
We choose this extreme situation of very shallow donor levels, for which EMT 
predictions for the binding energy work remarkably well, to explore the limits 
and capabilities of the TB formalism within a
recently proposed scheme \cite{form1}.
 
We consider a single impurity placed in a large cubic supercell containing $N=8
(L/{\rm a_{GaAs}})^{3}$ atoms arranged in the zincblende structure, where $L$ 
is the  supercell edge length. Periodic boundary conditions are imposed, and 
supercells containing up to 2 million atoms ($L = 64 \times{\rm a_{GaAs}}=361.6
$~\AA) are considered. We use the $sp^{3}s^{\ast}$ basis for the TB description 
of the electronic structure, with the bulk GaAs Hamiltonian taken from Boykin's 
parametrization \cite{boykin}, which includes first and second neighbor 
interactions and leads to an accurate value for the conduction band effective 
mass at $\Gamma$. 
Due to the $s$-character of the GaAs conduction band edge, spin-orbit 
corrections may be neglected in our calculations. The TB Hamiltonian is written 
as \cite{form1}
\begin{equation}
H=\sum\limits_{ij\mu\nu}h_{ij}^{\mu\nu}c_{i\mu}^{\dagger}c_{j\nu}+
\sum_{i\mu} U(r_{i}) c_{i\mu}^{\dagger}c_{i\mu} 
\label{hamilt} 
\end{equation}
where $i$ and $j$ denote the sites in the zincblende structure, $\mu$ and $\nu$ 
denote the atomic orbitals and $r_{i}$ is the distance from site $i$ to the 
impurity site. The $h_{ij}^{\mu\nu}$ define all the on-site energies and first 
and second neighbor hoppings for the bulk material. The perturbation potential 
$U(r_{i})$ is described by Eq.(\ref{pot}), except at the impurity site $(r_{i}
=0) $, where it is assigned the value $U_{0}$, describing central cell effects 
according to the substitutional species. In the present calculations, $U_{0}$ 
is kept as a free parameter. Estimates for this parameter are of the order of 
one to a few eV \cite{HVWD,form1}. Note that the impurity potential in (\ref
{hamilt}) incorporates both the
long-range component of the original EMT formulation \cite{emt2} and the short-
range perturbation ($U_0$ restricted to the impurity site) of the original TB 
formulation 
\cite{HVWD}. 

\begin{figure}
 \setlength{\unitlength}{1mm}
 \begin{picture}(150,150)(0,0)
 \put(0,0){\epsfxsize=10cm\epsfbox{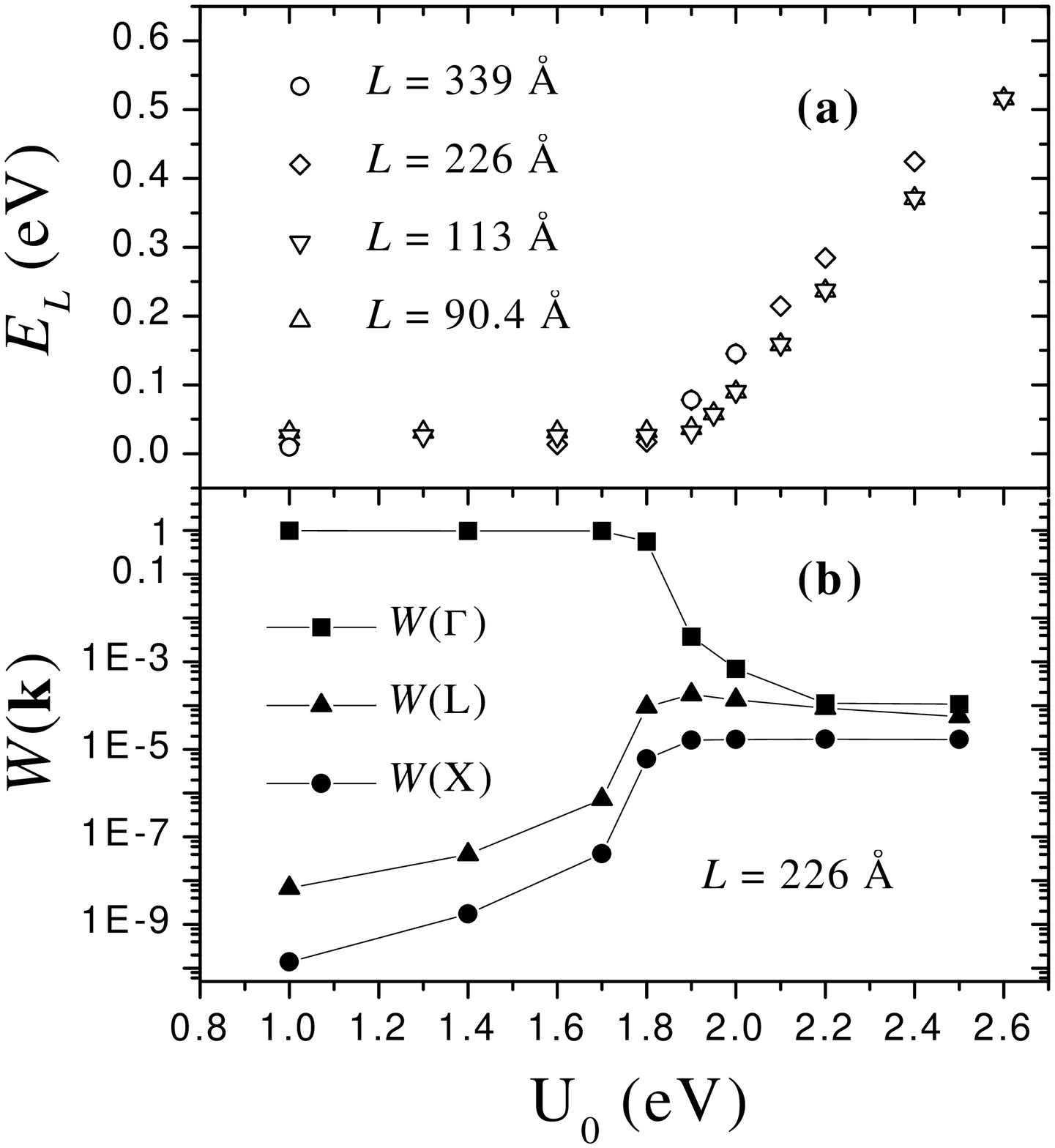}}
 \end{picture}
 \bigskip\bigskip
 \caption{(a) Binding energy of the impurity level as a function of the 
central cell perturbation potential $U_{0}$ for
 different supercell sizes $L$.  
(b) Calculated spectral weights for 
the $\Gamma$, X and L $k$-points as a function of $U_{0}$. 
The solid lines are guides for the eye. 
Both (a) and (b) revel a clear shallow-to-deep
level transition around $U_{0}=1.8$ eV. }
\label{fig:transi} 
\end{figure}

The exact ground state wave function and binding energy $E_{L}$ for a donor 
level within a supercell of size $L$ was obtained by  minimizing the 
expectation value of $\left\langle \Psi\left|  \left(H-\varepsilon_{ref}
\right)  ^{2}\right|  \Psi \right\rangle$, where $\varepsilon_{ref}$ is a 
reference energy chosen below the conduction band minimum \cite{araujo}, with 
the TB wave function expansion coefficients $\{C_{i\mu}\}$ taken as independent 
variational parameters. This procedure is equivalent to the exact 
diagonalization of $H$ for the eigenvalue and eigenfunction closest to 
$\varepsilon_{ref}$. The calculated binding energies as a function of $U_0$  
are given in Fig.~\ref{fig:transi}(a). It is clear that there is a range of the 
values of $U_{0}$ for which $E_{L}(U_0)$ is constant, indicating that the 
binding energy  does not depend on the impurity species, in agreement with the 
EMT and with experiments \cite{exper}. A well defined shallow-to-deep 
transition occurs for $U_{0}\approx 1.8$~eV, above which the binding energy 
increases approximately linearly with $U_0$. Aiming at a better 
characterization of this transition, we calculate the orbital averaged spectral 
weight of the donor state, 
 \begin{equation}
 W\left(  \bf{k}\right)  =\frac{2}{N}\sum\limits_{\mu
 ,i,j}e^{i\bf{k}\cdot(  {\bf R_{i}}-{\bf R_{j}})}C_{i\mu}C_{j\mu},
 \end{equation}
where ${\bf R}_{i}$ and ${\bf R}_{j}$ denote the position vectors of the
$i$ and $j$ atomic sites. 
The calculated $W\left(  \bf{k}\right)$ for ${\bf k} =\Gamma$, X and L
points in the
{\it fcc} Brillouin zone are given  in  Fig.~\ref{fig:transi}(b).
We visualize in this Figure the {\it k}-space counterpart  of the shallow-to-deep 
transition around $U_{0}=1.8$~eV. For $U_{0}\lesssim~1.8$~eV, the wave function 
has an almost pure $\Gamma$ character, since $W(\Gamma)$ is very close to 1. 
This extreme localization in $k$-space is consistent with the EMT assumptions, 
and indicates a highly delocalized wave function in real space. Increasing $U_
{0}$ beyond 1.8 eV, the dominant $\Gamma$ character breaks down and the impurity 
state must be described in terms of the several $\bf{k}$-point components, indicating delocalization in {\it k}-space, localization in real space. This is consistent 
with the deeper character of the state obtained in Fig.~\ref{fig:transi}(a) in 
this range of $U_0$.

In their original TB study, Hjalmarson {\it et al}~\cite{HVWD} noted that 
shallow levels are unbound by the central cell potential alone, and thus could 
not be obtained there. Our calculations are in complete agreement with this 
interpretation: We find that, in the whole range of $U_0$ for which $E_L (U_0)$ 
is constant, no bound state is obtained if the Coulomb potential is not 
included in the impurity perturbation. This result confirms a simple physical 
criterion characterizing shallow levels, and the reason why they do not depend 
on the substitutional impurity species, i.e. on $U_0$, since the binding is 
entirely due to the Coulomb attraction part of the perturbation potential. On 
the other hand, when deeper levels are obtained, $U_0$ alone also leads to a 
bound state, and therefore the binding energy is sensitive to the impurity 
species.

\begin{figure}
 \setlength{\unitlength}{1mm}
 \begin{picture}(150,150)(0,0)
 \put(0,0){\epsfxsize=10cm\epsfbox{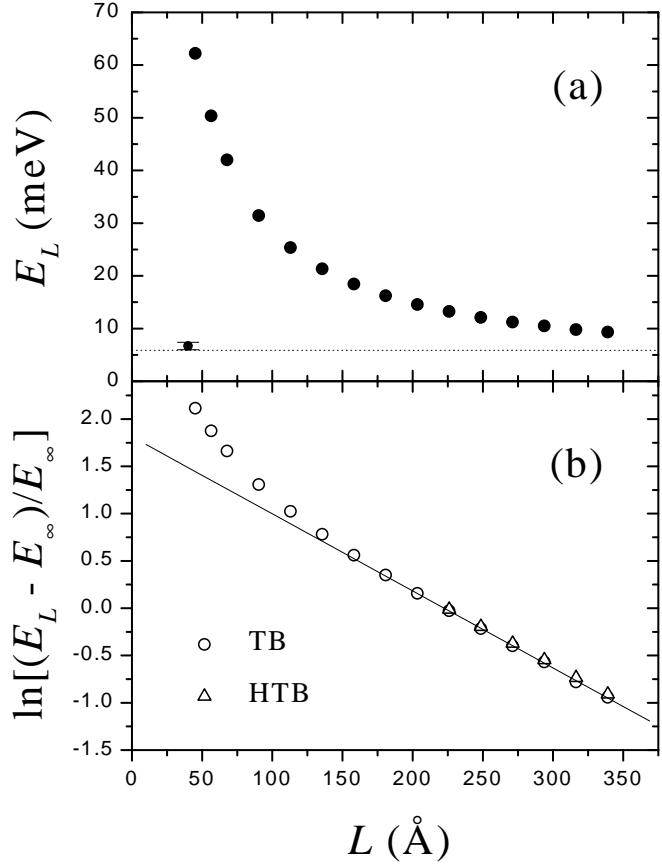}}
 \end{picture}
 \caption{(a) Tight-binding results for the binding 
energy of a single donor substitutional impurity in 
GaAs as function of supercell size $L$, with a
central cell correction $U_{0}=1.0$~eV. 
The dotted line represents the EMT value ($E_{d}^{*} =5.86$~meV), which
is in good agreement with experimental data. The circle at the lower left  gives our extrapolated value for $L\to \infty$, $E_{\infty} =6.7\pm{0.7}$~meV. (b) Exponential 
convergence of the donor binding energy for both TB and HTB calculations described in 
the text. The line is an exponential fit to the TB results (see text) with $E_{\infty} =6.7$~meV.}
 \label{fig:energy} 
 \end{figure}

In the following we focus on shallow donors. Given that the binding energy is 
independent of $U_0$ in this regime, we arbitrarily take $U_{0} = 1.0$ eV in 
the model calculations. This is a reasonable choice given the value of the 
Coulombic potential at the nearest-neighbor distance.
Fig.~\ref{fig:energy}(a) gives the donor ionization energy calculated at 
increasing supercell sizes. Results for $E_L$ presented there, even for the 
largest sizes, are still
decreasing with $L$.
This difficulty was overcome in Ref.\cite{form1} by assuming that,
asymptotically,  
the convergence of $E_L$ to the isolated-impurity  limit ($L
\rightarrow\infty$)
should be exponential, namely 
 \begin{equation}
 E_{L}=E_{\infty}+\widetilde{E}e^{-L/\lambda}.
 \label{extrapol}
  \end{equation}
 where  $\widetilde{E}$,  $\lambda$ and the donor ionization energy for
infinite cells
 $E_{\infty}$ are taken as adjustable parameters.  
The validity of Eq.(\ref{extrapol}) implies a linear dependence of
$\ln[(E_L-E_\infty)/E_\infty]$ on the system size $L$. 
These are the circles in Fig.~\ref{fig:energy}(b). 
As indicated by the straight solid line, 
a linear behavior is obtained for $L/a^* \gtrsim 1.5$  
with the extrapolated bulk donor binding energy $E_\infty = 6.7$ meV and 
$\lambda =1.25 a^*$.
We note a much slower convergence of the donor shallow level with $L$ as
compared to previous calculations for acceptors \cite{form1}. Similar fits were 
obtained for $E_{\infty}=6.7\pm0.7$ meV, while values of $E_\infty$ outside 
this range lead to systematic deviations of the calculated data points from a 
linear behavior. The extrapolated binding energy $E_{\infty}$ is indicated in 
Fig. \ref{fig:energy}(a). Although the EMT value is slightly off the lower 
limit estimated numerically, our results do not indicate any significant 
discrepancy between the binding energies determined by the two approaches
\cite{foot}.

\begin{figure}
 \setlength{\unitlength}{1mm}
 \begin{picture}(150,150)(0,0)
 \put(0,0){\epsfxsize=12cm\epsfbox{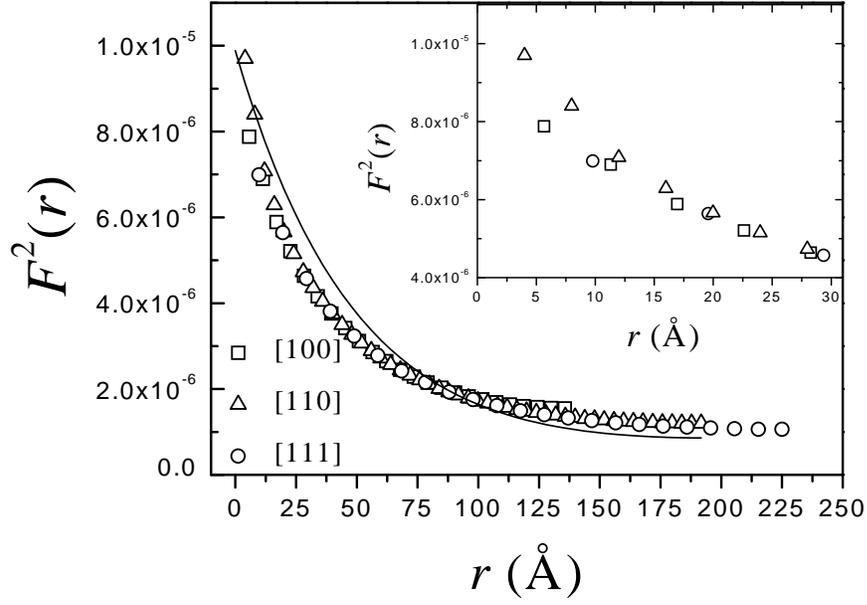}}
 \end{picture}
 \caption{TB envelope function squared along the indicated lattice directions. 
The solid line is a sum of exponential functions centered at each periodic 
replica of the
impurity, decaying with the EMT Bohr radius of 97.7 \AA.
The inset illustrates the anisotropy of the TB wave function around the
impurity site.}
 \label{fig:func} 
 \end{figure}

We now focus on the analysis of the donor wave function. Fig.~\ref{fig:func} 
gives the TB envelope function squared (expansion coefficients squared, summed 
over the five orbitals for the cation sublattice), calculated for $L=48\, {\rm 
a}_{\rm GaAs}\approx 270$\AA, as a function of distance from the impurity along 
the [100], [110] and [111] directions. At long distances, the wave function 
shows no angular dependence, as predicted by EMT. However, in the vicinity of 
the impurity there are noticeable anisotropies, which are highlighted in the 
inset. This is an effect of the crystalline environment that is automatically 
captured in an atomistic method such as TB. A more important discrepancy with 
respect to EMT is that {\em the donor-state wave functions cannot be described 
by a simple exponentially decaying behavior}. In fact, the line in Fig. \ref
{fig:func} is a sum of exponential functions centered at each supercell-periodic 
replica of the impurity, decaying with the EMT Bohr radius of 97.7 \AA. The 
agreement is not good, and it cannot be improved by varying the Bohr radius.

\begin{figure}
 \setlength{\unitlength}{1mm}
 \begin{picture}(150,150)(0,0)
 \put(0,0){\epsfxsize=12cm\epsfbox{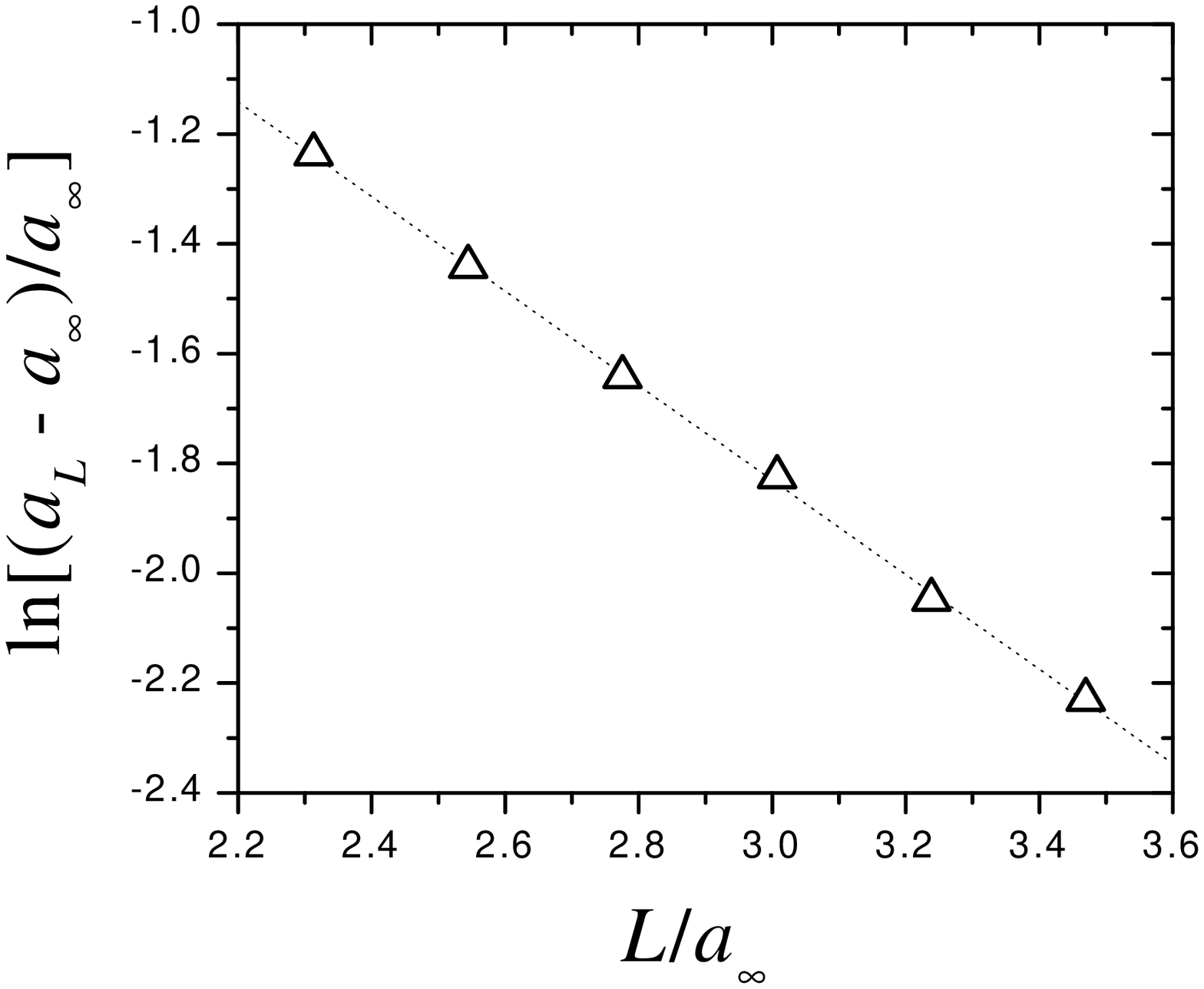}}
 \end{picture}
 \caption{Exponential convergence of the Bohr radius with the supercell size \textit{L} in the HTB model.  The dotted line corresponds to $a_\infty = 93$~\AA.}
 \label{fig:RBconv} 
 \end{figure}

It is not surprising that EMT works well for the binding energies but not so 
well for wave functions. This can be seen as a manifestation of the variational 
principle: Small variations in the ground-state wave functions do not affect 
the ground-state energies in linear order. We test this hypothesis through a 
set of variational calculations in which trial wave functions are constructed 
by rescaling the TB expansion coefficients of the true wave function in such a 
way that they are constrained to an envelope composed of summing exponential 
functions centered at each periodic replica of the impurity. In other words, in 
these test calculations, the TB wave function expansion coefficients $\{C_{i\mu}
\}$ are not independent variational parameters, but follow an exponential-decay 
behavior, corrected for the periodic boundary conditions. The decay length is 
now the sole variational parameter and it is then varied to minimize energy. We 
name these calculations ``hydrogenic tight-binding'' (HTB). The binding 
energies resulting from these constrained variational calculations also scale 
with supercell size as in Eq.~(\ref{extrapol}), as shown by the triangles in 
Fig.~\ref{fig:energy}(b). The extrapolated value, $E_\infty^{HTB} =  6.4 \pm 0.6
$ meV, is for all practical purposes indistinguishable from the full TB value. 
We also calculate, for different supercell sizes $L$, the optimum decay length 
$a_L$, as illustrated in Fig.~\ref{fig:RBconv}. We find that the calculated lengths also converge exponentially:
 \begin{equation}
 a_{L}=a_{\infty}+\widetilde{a}\,e^{-L/\lambda}~.
 \label{ansatz}
 \end{equation}
The extrapolated Bohr radius is $a_\infty = (93 \pm 5)$\AA, entirely consistent 
with the EMT prediction! Therefore, even though the true wave function is not 
exponentially decaying, if such a constraint is imposed (as in EMT), the 
optimum Bohr radius agrees with EMT predictions and the binding energy is 
essentially correct.

In conclusion, we study the problem of shallow donors in GaAs within the TB 
formalism. The use of large supercells (up to $L=361.6$~\AA), together with a 
finite-size scaling procedure, allows an unprecedented test of EMT for shallow 
levels in semiconductors. We find good agreement between TB and EMT in the 
determination of the binding energy, but hydrogenlike wave functions do not 
provide an accurate description of the impurity state. In addition, by varying 
the central cell corrections to the potential, we confirm the EMT predictions 
that the binding energy is independent on atomic impurity species for shallow 
levels. Our results are inconsistent with the commonly accepted idea that 
it is straightforward to describe bound states within atomistic approaches
if the supercell is large compared to the Bohr radius (or typical localization 
lengths). For the binding energy, the value obtained for a cell at least 3 times 
the EMT Bohr radius is about 30\% above the converged value. 
An extrapolation scheme is required to reach reliable results. 
Description of the wave function is even more subtle, due to the adopted boundary conditions.
Periodic boundary conditions requires careful consideration of contributions from 
the impurity periodic images. 
We demonstrate these inconsistencies and propose general and original ways to overcome 
them, applicable to any atomistic approach.

We thank H. Chacham and F.J. Ribeiro for helpful discussions, and CNPq, CAPES, 
FAPERJ, MCT-PRONEX, and FUJB for financial support.

\begin{thebibliography}{9}
\bibitem{foot} Note the large energy scales involved in the problem, e.g. in 
Figs.~\ref{fig:transi}(a) and ~\ref{fig:energy}(a), from which differences 
of tens of a meV may well be due to some degree of arbitraryness in the 
fitting procedures or small inaccuracies in our numerical variational scheme.
\end{thebibliography}
\end{document}